\documentclass[pra,aps,amsfonts,amsmath,superscriptaddress,twocolumn,showpacs,floatfix]{revtex4-1}
\usepackage{graphicx} 
\usepackage{epsfig} 
\usepackage{soul} 

\begin{document}

\title{Towards a better knowledge of the nuclear equation of state from the isoscalar breathing mode}

\author{E. Khan}
\affiliation{Institut de Physique Nucl\'eaire, Universit\'e Paris-Sud, IN2P3-CNRS, F-91406 Orsay Cedex, France}
\author{J. Margueron}
\affiliation{Institut de Physique Nucl\'eaire, Universit\'e Paris-Sud, IN2P3-CNRS, F-91406 Orsay Cedex, France}
\affiliation{Institut de Physique Nucl\'eaire de Lyon, Universit\'e de Lyon, IN2P3-CNRS, F-69622 Villeurbanne, France}

\begin{abstract}
The measurements of the isoscalar giant monopole resonance (GMR), 
also called the breathing mode, are analyzed with respect to their
constraints on the quantity $M_c$, e.g. the density dependence of
the nuclear incompressibility around the so-called crossing density
$\rho_c$=0.1 fm$^{-3}$. The correlation between the centroid of the
GMR, $E_\mathrm{GMR}$, and $M_c$ is shown to be more accurate than the
one between $E_\mathrm{GMR}$ and the incompressibility modulus at
saturation density, $K_\infty$, giving rise to an improved
determination on the nuclear equation of state. The relationship
between $M_c$ and $K_\infty$ is given as a function of the skewness parameter
$Q_\infty$ associated to the density dependence of the equation of
state. The large variation of $Q_\infty$ among
different energy density functionnals directly impacts the knowledge
of $K_\infty$: a better knowledge of $Q_\infty$ is required to deduce
more accurately $K_\infty$. Using the Local Density Approximation, a
simple and accurate expression relating $E_\mathrm{GMR}$ and the
quantity $M_c$ is derived and successfully compared to the fully
microscopic predictions.

\end{abstract}


\date{\today}

\maketitle

\section{Introduction}

The determination of the nuclear incompressibility is a long standing
problem. The earliest microscopic analysis came to a value of
$K_\infty$=210 MeV \cite{bla80}, but with the advent of microscopic
relativistic approaches, a value of $K_\infty$=260 MeV was obtained
\cite{vre03}. The fact that $K_\infty$ cannot be better determined
than 230$\pm$ 40 MeV, taking into account the whole data on the
isoscalar Giant Monopole Resonance (GMR) as well as the various
methods to relate the GMR to $K_\infty$ (see e.g.
\cite{bla80,vre03,agr03,col04,pie07,tli07,jli08,kha09,ves12}) lead to a recent
effort to reanalyse the method \cite{kha12}.

Pairing effects and similarly the shell structure effects on the
nuclear incompressibility were analyzed along these lines. Since the
first investigation~\cite{civ91}, several studies have shown that
pairing effects have an impact on the determination of
$K_\infty$~\cite{jli08,kha09}, and it was considered as a possible
cause of the difficulty to accurately constrain $K_\infty$.  This
effect of pairing on the incompressibility modulus has also been 
analyzed in nuclear matter, showing that the main effect is occurring
at subsaturation densities~\cite{kha10}.  However there is a general
consensus between the various microscopic models that pairing effects
on K$_\infty$ are not strong enough to explain the lack of accuracy in
the determination of the nuclear 
incompressibility~\cite{jli08,kha09,ves12,lig12}. Other effects have
to be investigated.

Recently, the density dependence of the nuclear incompressibility was
re-investigated suggesting that the correlation between the centroid
of the GMR and the incompressibility modulus $K_\infty$ at saturation
density is blurred by the density dependence of the nuclear equation
of state in different models~\cite{kha12}. The observed differences in
the extraction of $K_\infty$ from the $E_\mathrm{GMR}$ are based on
different models and attributed to the density dependence of the
equation of state (EoS) which has still to be better constrained. The
observation of a crossing point provided a possible path to be
investigated. The crossing point arises from Energy Density
Functionnals (EDFs) that are designed to describe finite-nuclei
observables: their density-dependent incompressibility $K(\rho)$
crosses around the mean density in nuclei,
$\rho_c\simeq0.1$~fm$^{-3}$. It was therefore proposed that the
quantity $M_c$, e.g. the density dependence of $K(\rho)$ around the
crossing density $\rho_c$ is the quantity that shall be constrained by
measurements of the GMR, instead of $K_\infty$.

The aim of the present article is to further analyze the correlation
method on which is extracted the incompressibility modulus $K_\infty$,
and to give a better basis on the alternative method based on the
correlation between $E_\mathrm{GMR}$ and $M_c$. A comparison between
the two methods is given in Sec.~\ref{sec:micro} for $^{208}$Pb and
$^{120}$Sn nuclei, showing the relevance of the new
method~\cite{kha12}. In Sec.~\ref{sec:dd} the source of uncertainties
in the determination of $K_\infty$ is directly related to the skewness
parameter $Q_\infty$. The skewness parameter gives indeed the present
limitation on the knowledge of the density dependence of the nuclear
EoS between the crossing and the saturation densities.  The origin of
the crossing density is also demonstrated in the case of the Skyrme
EDFs.  In Sec.~\ref{sec:lda}, the explicit relation between the
centroid of the GMR and the quantity $M_c$ is derived using the Local
Density approximation (LDA), and keeping as much as possible
analytical relations between the various quantities, in order to
facilitate their interpretation. The results are compared to the fully
microscopic one. Conclusions are given in Sec.~\ref{sec:conclusion}.

\section{The microscopic approach}
\label{sec:micro}

In this section, we first summarize the Constrained
Hartree-Fock-Bogoliubov approach (CHFB) used to accurately predict the
Isoscalar Giant Monopole Resonance (GMR) energy. We then provide the
definition of the parameter $M_c$, driving the density dependence of
the incompressibility around the crossing point. Finally, using these
two quantities, the correlation analysis between the GMR energy and
the incompressibility modulus $K_\infty$ on one hand, and the GMR
energy and the parameter $M_c$ at the crossing density on the other hand,
are compared.

\subsection{Microscopic calculation of the GMR energy}

We first recall how to predict the $E_{GMR}$. We use the sum rule approach in order to microscopically calculate the centroid energy
of the GMR. In such a microscopic approach, we calculate the energy as 
\begin{equation}
E_{\rm GMR}=\sqrt{\frac{m_1}{m_{-1}}}.
\label{eq:egmr}
\end{equation} 
where the $k$-th energy weighted sum rule is  defined as 
\begin{equation}
m_k=\sum_i(E_i)^k|\langle i|\hat{Q}|0\rangle |^2,
\end{equation}
with the RPA excitation energy  $E_i$  and the isoscalar monopole transition 
operator,
\begin{equation}
\hat{Q}=\sum_{i=1}^A r_i^2.
\end{equation}

The calculations using fully microscopic approaches based on EDF are
usually performed using CHFB or the RPA approach \cite{paa07}. In the
present case we calculate the GMR centroid for the Skyrme EDF with the
CHFB approach. For completeness, results using other functionnals such
as Gogny and relativistic functionnals will also be given. The CHFB
method is known to provide an accurate prediction of the GMR centroid.

In the following the energy weighted moment $m_1$ and the $m_{-1}$ moment are 
directly evaluated from the ground-state obtained from Skyrme CHFB calculations.
The moment $m_1$ is evaluated by the  double commutator using the 
Thouless theorem \cite{tho61}:
\begin{equation}
m_1=\frac{2\hbar^2A}{m} \langle r^2 \rangle. 
\end{equation}
where $A$ is the number of nucleons, $m$ the nucleon mass and 
$\langle r^2 \rangle$ is the rms radius evaluated on the ground-state
density given by Skyrme HFB.

Concerning the evaluation of the moment $m_{-1}$, the constrained-HFB
approach is used. It should be noted that the extension of the
constrained HF method \cite{boh79,col04} to the CHFB case has been
demonstrated in Ref. \cite{cap09} and employed in \cite{kha09}. The
CHFB Hamiltonian is built by adding the constraint associated with the
isoscalar monopole operator, namely 
\begin{equation}
\hat{H}_{constr.}=\hat{H}+\lambda\hat{Q}, 
\end{equation}
and the $m_{-1}$ moment is obtained from the derivative of the expectation value of 
the monopole operator on the CHFB solution $\vert\lambda\rangle$,
\begin{equation}
m_{-1}=-\frac{1}{2}\left[\frac{\partial}{\partial\lambda}\langle\lambda|
\hat{Q}|\lambda\rangle\right]_{\lambda=0}.
\end{equation}

\subsection{Constraints on the equation of state deduced from $E_\mathrm{GMR}$}

Next, the parameter $M_c$ is defined. Instead of correlating $E_\mathrm{GMR}$ 
and $K_\infty$, it was proposed that the energy of the GMR (\ref{eq:egmr}) gives a 
strong constraint on the quantity $M_c$ defined, at the crossing density $\rho_c\simeq$
0.1 fm$^{-3}$, as~\cite{kha12},
\begin{equation}
M_c\equiv 3\rho_c K'(\rho)|_{\rho=\rho_c} ,
\label{eq:edm}
\end{equation}
where the density-dependent incompressibility $K(\rho)$ is derived from the 
thermodynamical compressibility $\chi(\rho)$ as~\cite{fet71},
\begin{equation}
K(\rho)=\frac{9\rho}{ \chi(\rho)} = \frac{18}{\rho} P(\rho) +
9\rho^2\frac{\partial^2 E(\rho)/A}{\partial\rho^2},
\label{eq:krho}
\end{equation}
and the pressure is
\begin{equation}
P(\rho)\equiv\rho^2\frac{\partial E(\rho)/A}{\partial \rho}
\label{eq:press}
\end{equation}
The parameter $M_c$ was introduced instead of
$K_\infty\equiv K(\rho_0)$ (where $\rho_0$ is the saturation density)
in the correlation analysis based on $E_\mathrm{GMR}$ since i) the
crossing density $\rho_c$ is closer to the average density in finite
nuclei than the saturation density $\rho_0$, and ii) the crossing of
the incompressibility at $\rho_c$ makes the $E_\mathrm{GMR}$ mostly sensitive
to the derivative of the incompressibility at the crossing
density ~\cite{kha12}. It
should be noted that the existence of a crossing density for other EoS
quantities, such as for instance the symmetry energy \cite{pie11}, the
neutron EoS \cite{bro00,typ01} and the pairing gap in nuclear matter
\cite{kha09b}, was also observed. It might reveal the general trend
that the experimental constraints drive these quantities towards a
crossing point at around the average density of finite nuclei. Various
EDF's shall however exhibit various density dependencies around the
crossing point. At first order the derivative of the incompressibility
(or symmetry energy or pairing gap) at this point will differ between
various EDF's and additional measurements in nuclei shall characterize
these derivatives.  For instance, the derivative of the neutron EoS
around $\rho_c\simeq 0.11$~fm$^{-3}$ was found to be strongly
correlated to the neutron skin in $^{208}$Pb~\cite{bro00,typ01}, 
giving a strong support to improved experimental measurements of this
quantity~\cite{PIREX}.

\begin{figure}[tb]
\begin{center}
\scalebox{0.35}{\includegraphics{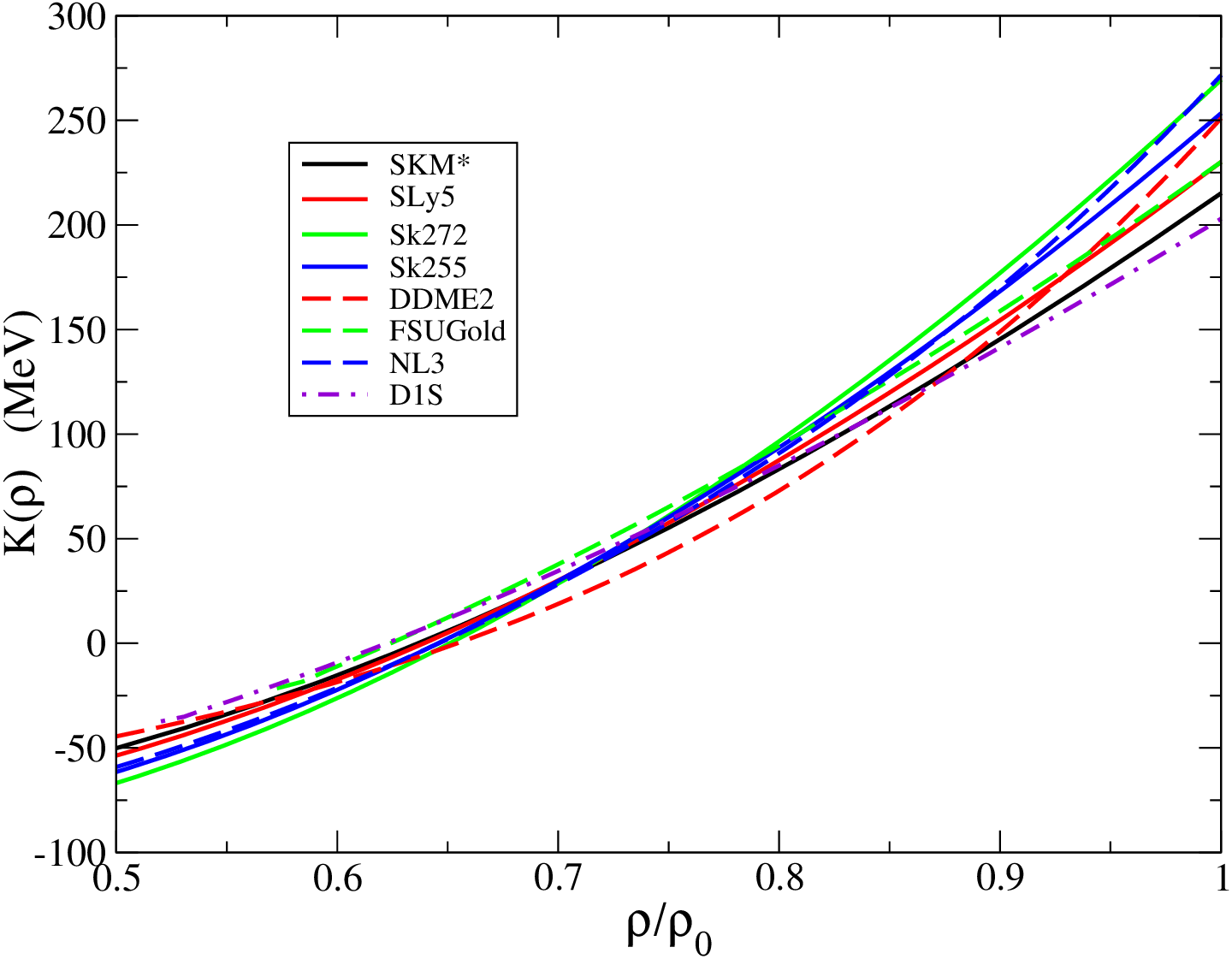}}
\caption{(Color online) EoS incompressibility $K(\rho)$ calculated with various
relativistic and non-relativistic functionnals.
The Skyrme EDF are in solid lines}
\label{fig:krho}
\end{center}
\end{figure}

Fig. \ref{fig:krho} depicts $K(\rho)$, between half of the saturation
density and the saturation density, for several Gogny, Skyrme and
relativistic EDFs.  A large dispersion
is observed at saturation density ($\rho/\rho_0=1$) whereas at
$\rho/\rho_0\simeq 0.71$ there is a much more focused area, defining
the crossing density $\rho_c$.

We now analyze the two contributions to the density-dependent
incompressibility, as depicted by Eq.~(\ref{eq:krho}). The first term
of the r.h.s in Eq.~(\ref{eq:krho}) is proportional to the pressure $P(\rho)$, which
is indeed related to the first derivative of $E/A$ and the second term is the
second derivative of the binding energy $E/A$ with respect to the
density. The former vanishes at saturation density, by definition.
Table ~\ref{tab:coefK} displays the expectation values of these two
contributions to the incompressibility $K(\rho)$~(Eq. (\ref{eq:krho}))
at the crossing density $\rho_c=0.71\rho_0$ and at the saturation
density for several EDFs. The total value of $K(\rho_c)\equiv K_c$ at the
crossing density is also displayed, while at $\rho_0$, 
$K_\infty=9\rho_0^2\frac{\partial^2 E(\rho)/A}{\partial
\rho^2}\vert_{\rho_0}$. The very weak dispersion of $K_c$ as a
function of the EDFs is striking, whereas the incompressibility
modulus at the saturation density $K_\infty$ is more scattered. At the
crossing density, the incompressibility $K_c$ is given as the sum of
the first and second derivatives of the energy per particle $E/A$,
which act in opposite signs, see Table ~\ref{tab:coefK}.  The
contribution of the pressure at $\rho_c$ is not negligible, at
variance with its contribution at $\rho_0$, and largely contributes to
the stabilisation of $K_c$. The correlations between $E_\mathrm{GMR}$
and the solely second derivative of $E/A$ at the saturation density
might not be the most appropriate one and the EDF-invariant property
of the crossing point ($\rho_c$,$K_c$) shall be useful.

\begin{table}[t]
\setlength{\tabcolsep}{.02in}
\renewcommand{\arraystretch}{1.4}
  \begin{center}
  \begin{tabular}{c|ccc||ccc}
    \toprule
   
   & \multicolumn{3}{c||}{Crossing}&\multicolumn{3}{c}{Saturation}\\
   & 
   $K_c$ &
    $\frac{18}{\rho_c} P(\rho_c)$ & 
    $9\rho_c^2\frac{\partial^2 E(\rho)/A}{\partial \rho^2}\vert_{\rho_c}$ & 
    $K_\infty$ & $P(\rho_0)$ &
    $9\rho_0^2\frac{\partial^2 E(\rho)/A}{\partial \rho^2}\vert_{\rho_0}$ \\
   &   MeV   &   MeV   &   MeV   &   MeV  & MeV  & MeV\\
    \colrule
SLy5       & 36 &  -103 &139 & 230 & 0 & 230  \\
SkM$^*$& 34 &    -99   & 133 & 217 & 0   & 217  \\
Sk255     & 36 & -113 & 149 & 255 & 0 & 255  \\
Sk272     & 35 & -119 & 154 & 272 & 0 & 272  \\
SGII         &   34 & -98 & 132 & 215 & 0 & 215  \\
    \botrule
  \end{tabular}
  \end{center}
  \caption{Evaluation of $K_c$ and of the two terms defining 
  the incompressibility K($\rho$)~(Eq. (\ref{eq:krho})), at the
crossing density $\rho_c=0.71\rho_0$ and at the saturation density
$\rho_0$ for a set of different Skyrme EDFs.}
  \label{tab:coefK}
\end{table}

\subsection{($E_\mathrm{GMR}$,$K_\infty$) versus ($E_\mathrm{GMR}$,$M_c$) correlation analysis}

\begin{figure}[t]
\begin{center}
\scalebox{0.35}{\includegraphics{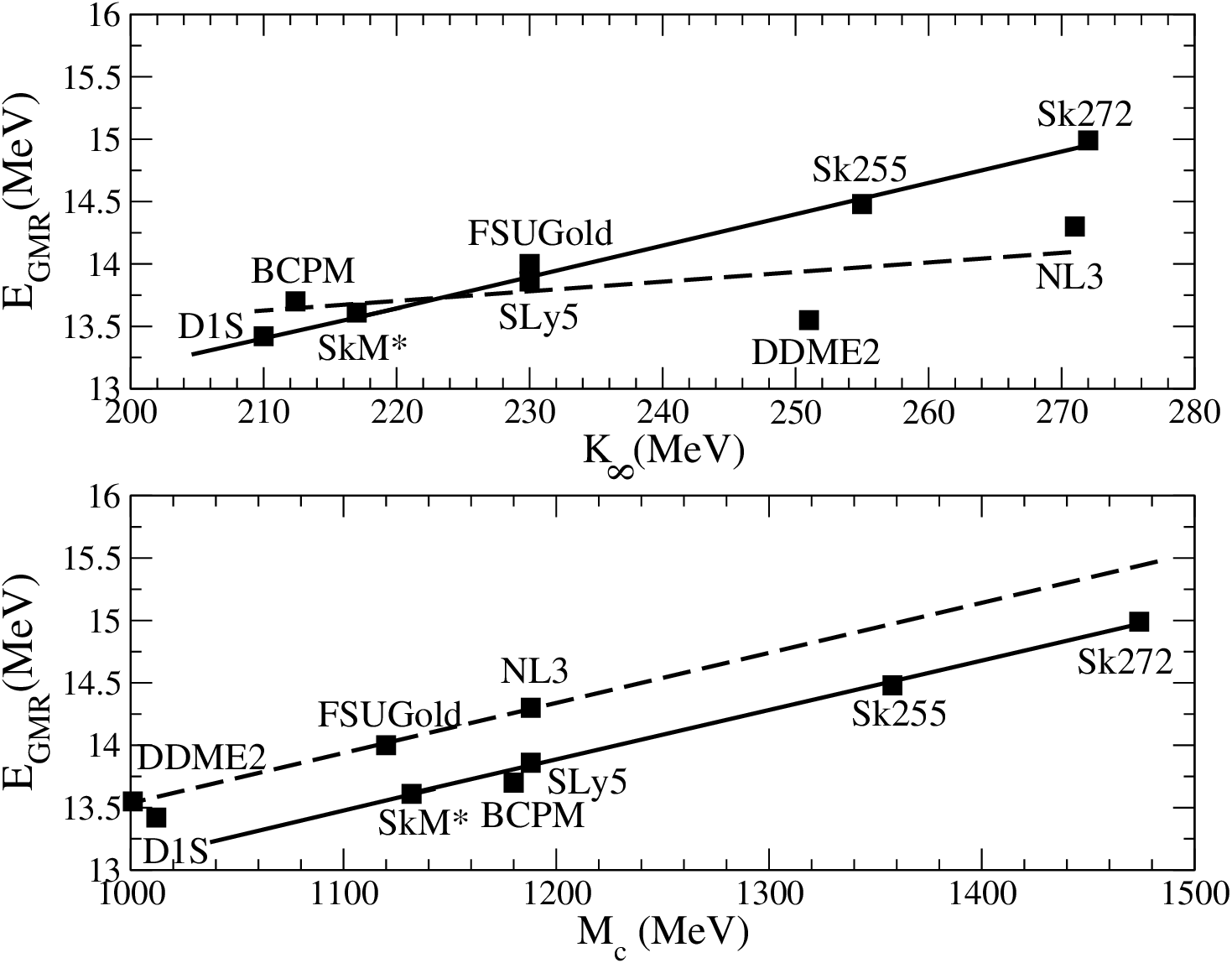}}
\caption{Centroid of the GMR in $^{208}$Pb calculated with the
microscopic method  (see text) vs. 
the value of $K_\infty$ (top) and  $M_c$ (bottom) for various functionals
\cite{agr03,pie07,lal97,lal05,bar82,ber91,cha98,kha12,bal12}.
The solid and dashed lines correspond to fit on the Skyrme and
relativistic EDF values, respectively.}
\label{fig:EMK}
\end{center}
\end{figure}

Using $E_\mathrm{GMR}$ and $M_c$ discussed in the previous sections,
it is possible to determine if $M_c$ is better constrained by
$E_\mathrm{GMR}$ than $K_\infty$. The correlation diagrams
($E_\mathrm{GMR}$,$K_\infty$) and ($E_\mathrm{GMR}$,$M_c$) are
compared on Fig. \ref{fig:EMK} in the case of $^{208}$Pb. The
relativistic DDME2 interaction in the correlation graph
($E_\mathrm{GMR}$,$K_\infty$) is largely deviating from the others, as
it is well known~\cite{vre03,agr03,col04} while it is much more
compatible with the others in the ($E_\mathrm{GMR}$,$M_c$) correlation
graph~\cite{kha12}. On the contrary, restricting to the Skyrme
interactions, the quantities ($E_\mathrm{GMR}$,$K_\infty$) and
($E_\mathrm{GMR}$,$M_c$) are equally well correlated. This is directly
related to the good correlation between ($M_c$,$K_\infty$) due to
a similar density dependence among the Skyrme EDFs (in $\rho^\alpha$),
which will be discussed in section III.C. However, considering various
models with different density dependencies, a better correlation is
observed between $E_\mathrm{GMR}$ and $M_c$, compared to the one
between $E_\mathrm{GMR}$ and $K_\infty$. It should be noted that in
the ($E_\mathrm{GMR}$,$M_c$) correlation graph, an ordering between
Skyrme and relativistic models is also observed.

\begin{figure}[t]
\begin{center}
\scalebox{0.35}{\includegraphics{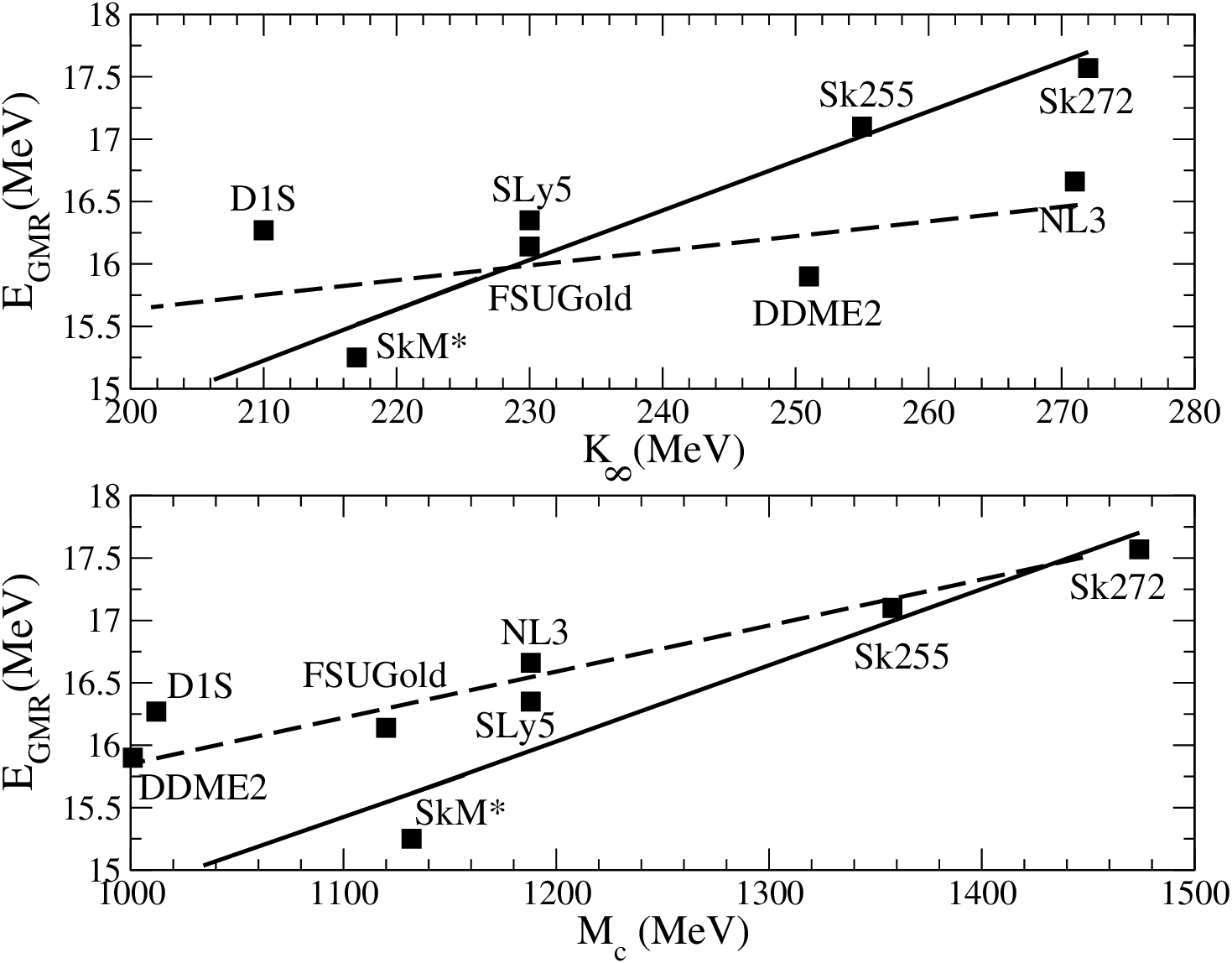}}
\caption{Same as Fig. \ref{fig:EMK} for $^{120}$Sn.}
\label{fig:EMKSn}
\end{center}
\end{figure}

Fig. \ref{fig:EMKSn} displays the same correlations in the case of the
$^{120}$Sn nucleus. In this case pairing effects are known to slightly
impact the position of the GMR \cite{kha10}, leading to a larger
dispersion compared to the $^{208}$Pb case. However similar
conclusions can be drawn, namely i) $M_c$ is a better correlated
quantity with $E_\mathrm{GMR}$ than $K_\infty$, ii) a good
($E_\mathrm{GMR}$,$K_\infty$) correlation is also observed among the
Skyrme interactions and iii) the ($E_\mathrm{GMR}$,$M_c$) correlation
exhibits an ordered correlation between the Skyrme and the
relativistic EDFs.

To conclude, these results on $M_c$ provide a
step towards compatible results between Skyrme, Gogny and relativistic
approaches~\cite{kha12}. The extracted value for the quantity $M_c$ in $^{120}$Sn and $^{208}$Pb 
nuclei are also in better agreement between each other than the corresponding
$K_\infty$ values: considering the various EDFs as well as the $^{120}$Sn and
the $^{208}$Pb data, one gets $M_c$=1100 $\pm$ 70 MeV (6\% uncertainty),
and $K_\infty$=230 $\pm$ 40 MeV (17\% uncertainty) \cite{kha12}. 
It should be noted that these considerations on the slope of the
incompressibility $M_c$ at the crossing point have recently been used in
Ref. \cite{bal12} where a good linear correlation between $E_\mathrm{GMR}$
and $M_c$ is also found, including the so-called BCPM functionnal.

In summary, using microscopic approaches, it is observed that the
correlation between $M_c$ and the centroid $E_\mathrm{GMR}$ is less
dispersive, and therefore more universal among various models,
than the one between $K_\infty$ and $E_\mathrm{GMR}$~\cite{kha12}.  In
the next section, we shall provide a more quantitative understanding of
the differences between the quantities $M_c$ and $K_\infty$, explaining
the role of the density dependence of the equation of state between
the crossing and the saturation densities.

\section{Density expansion of the equation of state}
\label{sec:dd}

The striking stability of $K_c$ among the various Skyrme EDFs (Table
~\ref{tab:coefK}) deserves an investigation. In this section, the
density dependence of the equation of state is discussed in terms of
the derivatives of the EoS with respect to the density. The stability
of $K_c$ as well as the relation between the slope of the 
incompressibility modulus $M_c$ and the parameters $K_\infty$ and
$Q_\infty$ are derived, providing an explanation for the difficulty to
constrain $K_\infty$.

\subsection{Density dependence of the equation of state
around $\rho_0$}

We start from a systematic expansion around the saturation density $\rho_0$ 
of the binding energy, such as in the Generalized Liquid Drop Model 
(GLDM)~\cite{Ducoin2010,Ducoin2011}
where, in symmetric matter, the energy per particle reads
\begin{equation}
E(x)=E_\infty+\frac{1}{2}K_\infty x^2+\frac{1}{6}Q_\infty x^3\dots
\label{eq:gldm}
\end{equation}
with $x=(\rho-\rho_0)/(3\rho_0)$, $\rho_0$ being the saturation density
of symmetric nuclear matter. $Q_\infty$ is the third derivative of the energy per particle.

Applying Eqs.~(\ref{eq:krho}) and (\ref{eq:press}) to the expansion Eq. ~(\ref{eq:gldm}), 
one obtains the pressure,
\begin{equation}
P(x)=\frac{1}{3}(1+3x)^2\Big[K_\infty x+\frac{1}{2}Q_\infty x^2+\dots\Big],
\label{eq:gldmp}
\end{equation}
and the incompressibility,
\begin{eqnarray}
K(x)&=&(1+3x)\Big[K_\infty +(9K_\infty+Q_\infty) x+6Q_\infty x^2 \nonumber \\
&&\hspace{2cm}+\dots\Big].
\label{eq:gldmk}
\end{eqnarray}

\begin{figure}[t]
\begin{center}
\scalebox{0.5}{\includegraphics{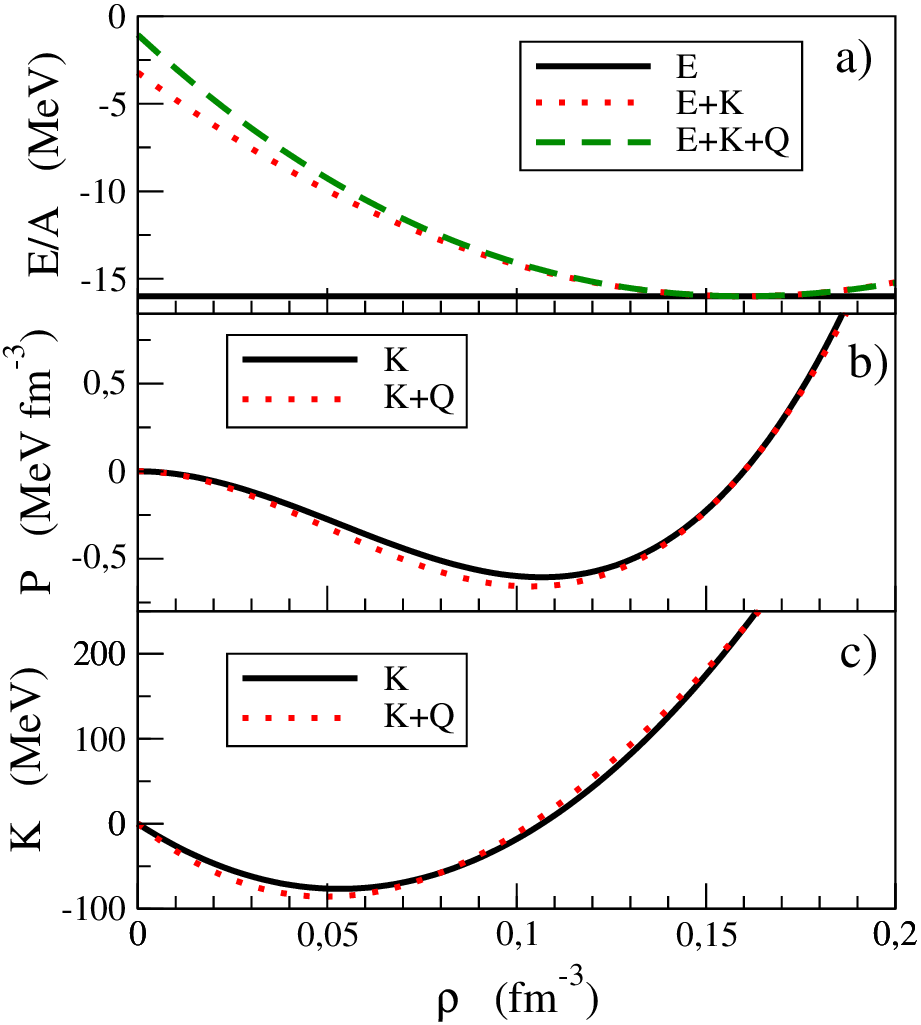}}
\caption{(color online) a) Binding energy $E/A$ in MeV, b) pressure in
MeV~fm$^{-3}$ and c) 
incompressibility K in MeV, as a function of the density for various truncation in the expansion 
Eq. (\ref{eq:gldm}). See text for more details.}
\label{fig:eos}
\end{center}
\end{figure}

Fig.~\ref{fig:eos} displays the binding energy Eq. (\ref{eq:gldm}),
pressure Eq. (\ref{eq:gldmp}) and incompressibility Eq.
(\ref{eq:gldmk}) as function of the density $\rho$ going from 0 to 0.2
fm$^{-3}$, for typical values for the quantities :
$E_\infty$=-16~MeV, $K_\infty$=240~MeV and $Q_\infty$=-350~MeV. The
different curves correspond to various approximations in the density
expansion of the binding energy Eq. (\ref{eq:gldm}). For instance, the
solid line in the binding energy $E/A$ corresponds to the $0$-th order
in the density expansion Eq. (\ref{eq:gldm}) where only the
quantity $E_\infty$ has been included, all other quantities being
set to zero. The dotted-line (E+K) takes into account the quantities
$E_\infty$ and $K_\infty$, and the dashed line (E+K+Q) includes the
quantities $E_\infty$, $K_\infty$ and $Q_\infty$. Similar
approximations have been performed in the case of the pressure Eq.
(\ref{eq:gldmp}) and incompressibility Eq. (\ref{eq:gldmk}). A good
convergence is found when successively including in the expressions
for the binding energy, the pressure and the incompressibility the
quantities $E_\infty$, $K_\infty$ and $Q_\infty$. These quantities
describe the density dependence of the equation of state and are given
in table~\ref{tab:coef} for a set of models considered in this work.

\begin{table}
\setlength{\tabcolsep}{.12in}
\renewcommand{\arraystretch}{1.5}
  \begin{center}
  \begin{tabular}{cccccc}
    \toprule
     & $\rho_0$ & $E_\infty$ & $K_\infty$ & $Q_\infty$  \\
   & fm$^{-3}$ &  MeV   &   MeV   &   MeV   \\
    \colrule
SLy5       & 0.160 & -15.98 & 230 & -363  \\
SkM$^*$& 0.160 & -15.79 & 217 & -386  \\
Sk255     & 0.157 & -16.35 & 255  & -350  \\
Sk272     & 0.155 &  -16.29 & 272  & -306  \\
D1S       & 0.163  & -16.02 & 210 & -596  \\
NL3         & 0.148 & -16.24 & 271 & 189  \\
DDME2   & 0.152 & -16.14 & 251 & 478  \\
FSUGold & 0.148 & -16.30 & 229 & -537  \\
    \botrule
  \end{tabular}
  \end{center}
  \caption{Parameters appearing in the density expansion of the binding energy
$E/A$  Eq. (\ref{eq:gldm}) for a set of models considered in this work:
$\rho_0$ is the saturation density, $E_\infty$ the binding energy, $K_\infty$ the 
incompressibility modulus, and $Q_\infty$ the skewness parameter.}
  \label{tab:coef}
\end{table}

It is clear from Table~\ref{tab:coef} that while the quantities
$E_\infty$ and $K_\infty$ are not varying by more than 20\%, the
values for the skewness parameter $Q_\infty$ is almost unconstrained
and can vary by more than 100\% among the models. The uncertainty in
the determination of the skewness parameter $Q_\infty$ gives, in a
quantitative way, the main lack of knowledge in the density dependence of the
equation of state. The uncertainty on $Q_\infty$ also impacts the
density dependence of the pressure and more, interestingly here, of
the incompressibility.

\subsection{Stability of $K_c$}

Let us now provide an explanation for the stability of $K_c$ observed
in Table~\ref{tab:coef}.
From Eq.~(\ref{eq:gldmk}), and assuming the validity of a density
expansion from $\rho_0$ to $\rho_c$, we obtain
\begin{equation}
K_c\simeq(1+3x_c)\left[(1+9x_c)K_\infty+(1+6x_c)x_cQ_\infty\right],
\label{eq:gldmkc}
\end{equation}
with $x_c=(\rho_c-\rho_0)/(3\rho_0)$.

\begin{figure}[tb]
\begin{center}
\scalebox{0.35}{\includegraphics{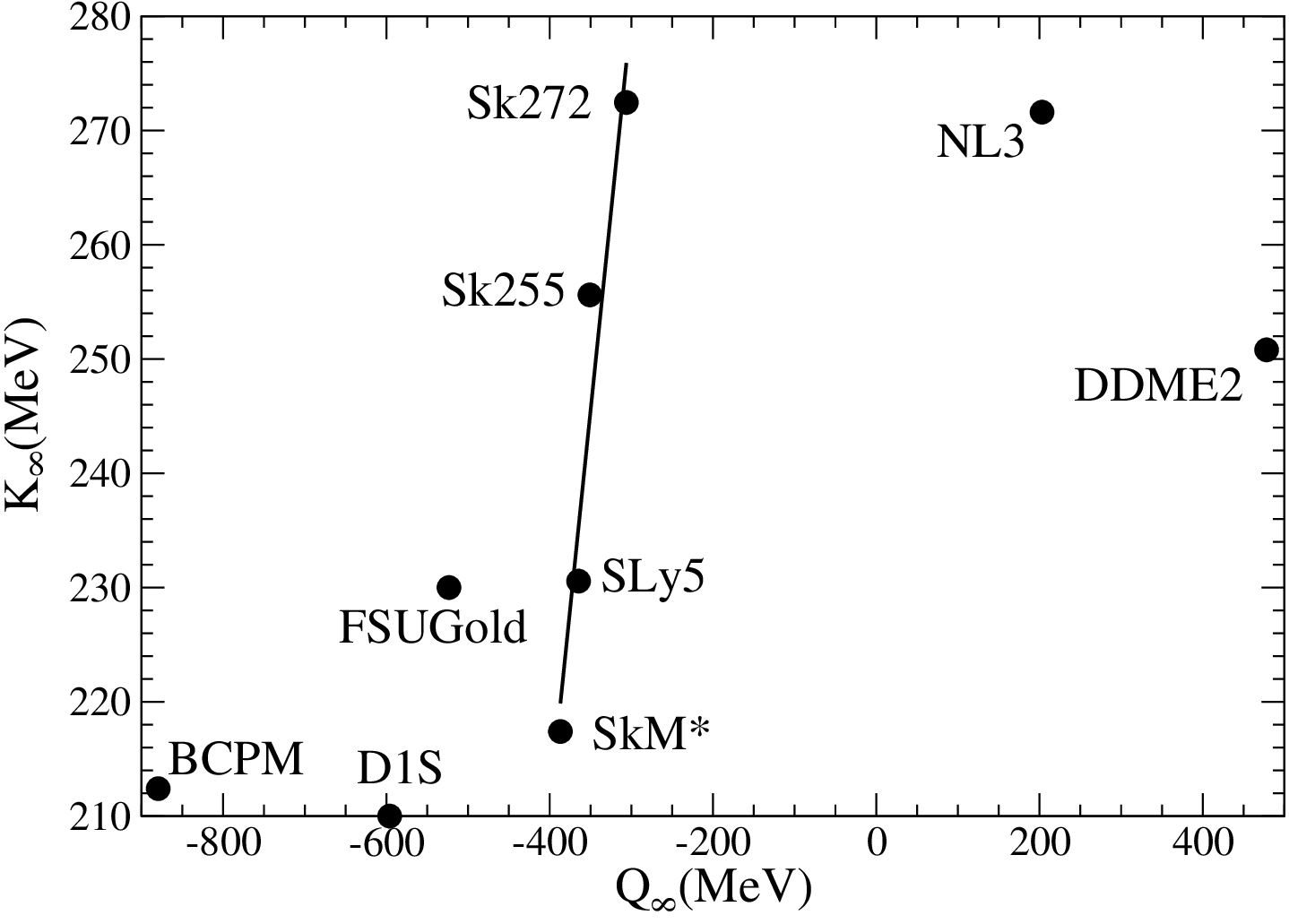}}
\caption{$K_\infty$ versus $Q_\infty$ for several models. The solid
line corresponds to the fit on the Skyrme EDFs values.}
\label{fig:KQ}
\end{center}
\end{figure}

In the case of Skyrme interaction, there is a good correlation among
the quantities $K_\infty$ and $Q_\infty$, as shown in
Fig.~\ref{fig:KQ}. The parameters $K_\infty$ and $Q_\infty$ are mostly
determined by the same term, the term $t_3$ in $\rho^\alpha$, in the
case of Skyrme interaction~\cite{kha12}. Due to their similar density 
dependence (in $\rho^\alpha$), the Skyrme 
EDFs exhibit indeed a linear correlation among these two quantities whereas the picture is
blurred when considering at the same time Skyrme and relativistic
EDFs. The linear correlation among the Skyrme EDFs can be described
by,
\begin{equation}
K_\infty=a+bQ_\infty ,
\label{eq:linearKQ}
\end{equation}
with $a = 338\pm 9$~MeV and $b=0.29\pm0.03$.
Injecting~(\ref{eq:linearKQ}) in Eq. (\ref{eq:gldmkc}), one gets
\begin{equation}
K_c\simeq(1+3x_c)\left[(1+9x_c)a+f(x_c)Q_\infty\right],
\label{eq:linkap}
\end{equation}
with $f(x)=(6x^2+(9b+1)x+b)$. An EDF-almost-independent value of $K_c$ is
therefore obtained for f(x)=0 since $Q_\infty$ is the only EDF-dependent
quantity in Eq. (\ref{eq:linkap}): the solution of f(x)=0 
shall therefore provide the crossing point observed on Fig. \ref{fig:krho}.
The function $f(x)$ has only one zero for positive densities, given by
$x_c=-0.095\pm0.002$, which corresponds to
$\rho=(0.714\pm0.005)\rho_0 = \rho_c$.

In the Skyrme case there is therefore a density, $\rho_c$, for
which the incompressibility modulus $K(\rho_c)$ is independent from the quantities
$K_\infty$ and $Q_\infty$ defined in Eq.~(\ref{eq:gldm}), and is
\begin{equation}
K_c\equiv K(\rho_c) = (1+3x_c)(1+9x_c)a =
\frac{\rho_c}{\rho_0}(3\frac{\rho_c}{\rho_0}-2)a .
\end{equation}
Taking the value for $a$ given by the linear correlation, one finds
$K_c=34\pm 4$~MeV, confirming the value of the crossing point
shown in Table~\ref{tab:coefK} and on Fig. \ref{fig:krho} for the
Skyrme EDFs. This approach confirms in a both quantitative and
qualitative way the existence of a crossing point, especially in the
case of the Skyrme EDF. In the case of the other EDFs, it is rather a
crossing area that is obtained (Fig. \ref{fig:krho}), due to their
various density dependence, as discussed above.

\subsection{Relation between $K_\infty$ and $M_c$}

\begin{figure}[tb]
\begin{center}
\scalebox{0.35}{\includegraphics{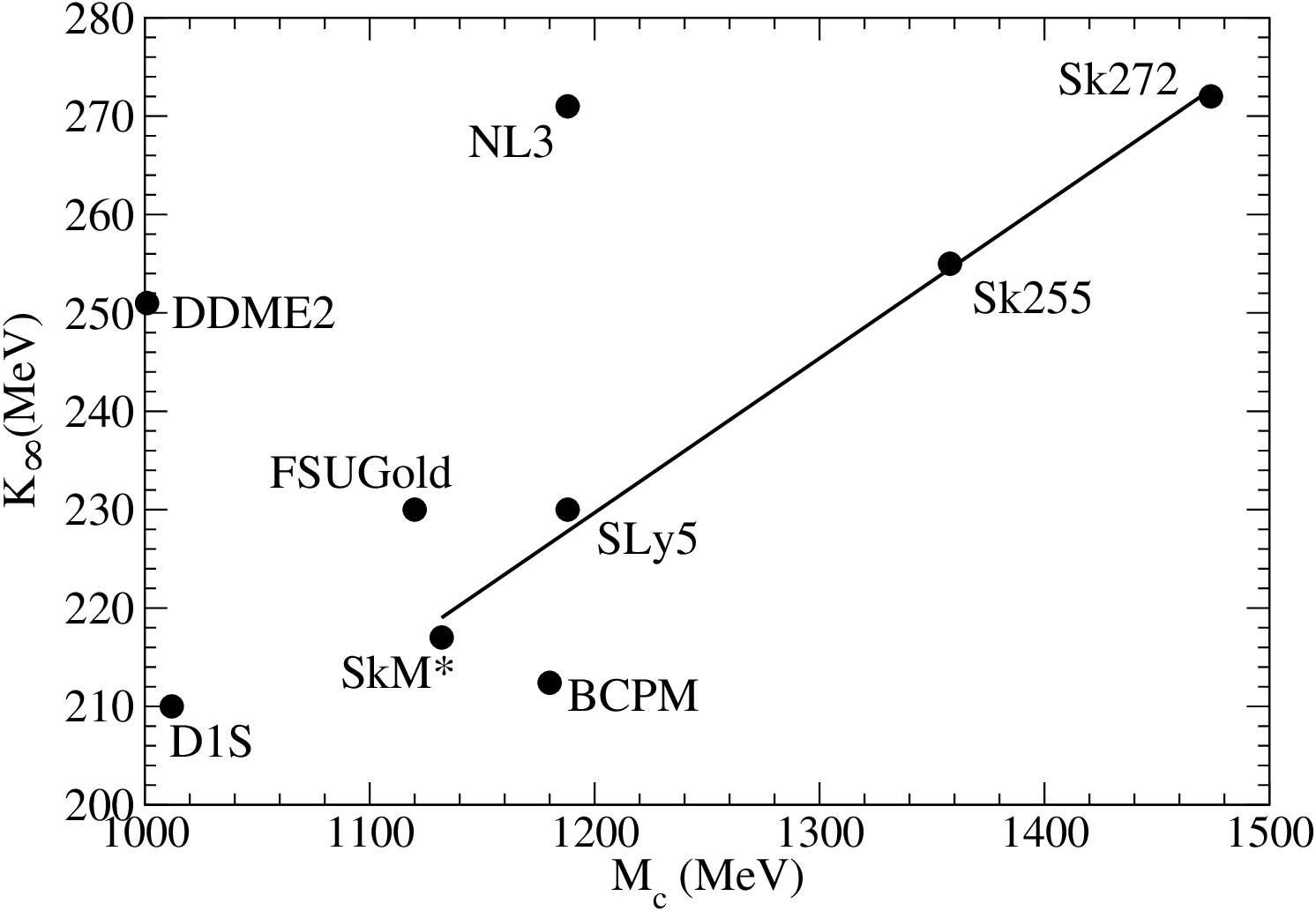}}
\caption{K$_\infty$ as a function of M$_c$ for various Skyrme,
Gogny and relativistic functionals}
\label{fig:kinfM}
\end{center}
\end{figure}

Fig. \ref{fig:kinfM} displays the ($M_c$,$K_\infty$) correlation
for four Skyrme EDFs. Adding to this correlation graph EDFs
with other density dependences, such as the relativistic one,
drastically blurs this correlation. This shall originates from the
large uncertainty on the value
of the skewness parameter $Q_\infty$
among the EDFs discussed in the previous section (Table~\ref{tab:coef}).
Using Eqs.~(\ref{eq:edm}), (\ref{eq:krho}) and (\ref{eq:gldmk}), the quantity $M_c$ 
can be expressed as,
\begin{eqnarray}
M_c &\simeq& 3K_c+(1+3x_c)^2\Big[9K_\infty+(1+12x_c)Q_\infty\Big] .
\label{eq:relM}
\end{eqnarray}
The correlation between $M_c$ and $K_\infty$ depends on the density dependence 
of the binding energy reflected in the skewness parameter $Q_\infty$, which can vary to a large 
extent, see Table~\ref{tab:coef}. More precisely, from Eq.~(\ref{eq:relM}), 
one can deduce the value of the quantity $K_\infty$ as,
\begin{equation}
K_\infty = \frac{1}{9}\frac{M_c-3K_c}{(1+3x_c)^2}-\frac{1+12x_c}{9}Q_\infty,
\label{eq:kk}
\end{equation}
where the second term of the r.h.s shows the theoretical error on
$K_\infty$ induced by the uncertainty on $Q_\infty$, the unconstrained
density dependence of the EoS. $K_c$ and $x_c$ are fixed by the
existence of a crossing point, and $M_c$ is extracted from the
correlation analysis based on the experimental $E_\mathrm{GMR}$. It is
therefore clear that the uncertainty on $K_\infty$ is related to the
lack of knowledge on the density dependence of the equation of state,
represented in the present analysis by the skewness parameter $Q_\infty$.
Taking a typical uncertainty for $Q_\infty$ of $\pm 400$~MeV
(Table~\ref{tab:coef}), Eq.
(\ref{eq:kk}) provides a variation on $K_\infty$ about $\pm 40$~MeV,
compatible with the present uncertainty on $K_\infty$.

In conclusion, the relation Eq. ~(\ref{eq:kk}) clearly shows that
the uncertainty on the incompressibility modulus $K_\infty$ is mainly
related to that on the quantities $Q_\infty$. The reduction of the
error bar on $K_\infty$ is therefore mostly related to a better
knowledge of the skewness parameter $Q_\infty$,
for which new experimental constraints shall be found. 
On the contrary the density dependence around the crossing point 
M$_c$ can be more directly constrained from $E_\mathrm{GMR}$, 
as it will be showed below.

\section{A simple expression relating $E_\mathrm{GMR}$ and $M_c$}
\label{sec:lda}

To provide a complementary view to microscopic approaches
\cite{bla80,vre03,agr03,col04,pie07,tli07,jli08,kha09,ves12,kha12}, it
may be useful to derive an analytic relationship between the GMR
centroid in nuclei and the quantity $M_c$, in order to enlighten and
confirm the results obtained with a fully microscopic approach, see 
Sec.~\ref{sec:micro} and Ref. \cite{kha12}. 
In this section we aim to derive an analytical relationship between
the centroid of the GMR and the relevant quantity of the EoS, $M_c$ Eq.~(\ref{eq:edm}).

The energy centroid of the GMR is used to define the incompressibility in nuclei
$K_A$ \cite{bla80}:
\begin{equation}
E_{\rm GMR}=\sqrt{\frac{\hbar^2K_A}{m \langle r^2 \rangle}} .
\end{equation}
In order to derive an analytical
relationship, $\langle r^2 \rangle$ can be approximated by $3R^2/5$
\cite{rin80} where $R\simeq1.2A^{1/3}$ is the nuclear radius, yielding:
\begin{equation}
E_{\rm GMR}\simeq\frac{\hbar}{R}\sqrt{\frac{5K_A}{3m}}
\label{eq:egmrka}
\end{equation}

We shall derive an analytic relation between $K_A$ and $M_c$ using the
LDA, in order to check, in a complementary way to microscopic
approaches, the role of $M_c$ in determining the centroid of the GMR.

The following step consists in dividing
$K_A$ into a nuclear and a Coulomb contributions, as
\begin{equation}
K_A=K_\mathrm{Nucl}+K_\mathrm{Coul}\cdot Z^2 A^{-4/3}
\label{eq:kaexp}
\end{equation}
where, in the liquid drop approach, $K_\mathrm{Nucl}$ is defined as
\begin{equation}
K_\mathrm{Nucl}=K_\infty+K_\mathrm{surf}A^{-1/3}+K_\tau\left(\frac{N-Z}{A}\right)^2 ,
\nonumber
\end{equation}
as in the Bethe Weiss\"acker formula for the binding
energy~\cite{bla80}. The accuracy of this approach can be enhanced
with the inclusion of higher order terms~\cite{pat02}. The
quantities $K_\infty$, $K_\mathrm{surf}$ and $K_\tau$ are however
poorly constrained by the relative small
data~\cite{bla80,Pearson2010}. We prefer instead to extract
$K_\mathrm{Nucl}$ from the LDA which has the advantage that i) it was
proven to be a good approximation of the microscopic
calculation~\cite{kha10}, and ii) the consistency between the value
obtained for $K_A$ and the Skyrme functional is guarantied.

\subsection{The local density approximation (LDA)}
 
The nuclear contribution $K_\textrm{Nucl}$ is related to the density dependence of the
incompressibility K($\rho$) as \cite{kha10},
\begin{equation}
K_\mathrm{Nucl}=\frac{\rho_0^2}{A}\int d^3r \; \frac{K(\rho(r))}{\rho(r)}
\label{eq:knucldef}
\end{equation}

Eq. (\ref{eq:knucldef}) allows to perform the LDA by considering the
density profile of nuclei, $\rho_A(r)$, in Eq. (\ref{eq:krho}), where
$\rho=\rho_A(r)$. The LDA give accurate estimation of $K_\mathrm{Nucl}$~\cite{kha10}.
It should be noted that in Eq.
(\ref{eq:knucldef}), the value of $K(\rho)$ at saturation density
(i.e. $K_\infty$) doesn't have any specific impact on the $K_A$ value
and therefore nor on the prediction of $E_\mathrm{GMR}$. Further, due
to the existence of the crossing area ($\rho_c$,$K_c$)$\simeq$(0.1
fm$^{-3}$,40 MeV), which takes into account both the Skyrme EDFs crossing
(Table \ref{tab:coefK}) and the relativistic one (Fig \ref{fig:krho}), 
$K(\rho)$ can be approximated to the first order around the crossing point by:
\begin{equation}
K(\rho)=\frac{M_c\rho}{3\rho_c}-\frac{M_c}{3} + K_c
\label{eq:mlin}
\end{equation}
where $M_c$ is related to the first derivative of the incompressibility,
Eq. (\ref{eq:edm}). 

This first order approximation is relevant as observed on the
($E_\mathrm{GMR}$,$M_c$) correlation of Figs. \ref{fig:EMK} and
\ref{fig:EMKSn}. Of course taking the density dependence of the
incompressibility as its first derivative around the crossing point
remains an approximation, which explains the non exactly linear
($E_\mathrm{GMR}$,$M_c$) correlation on Fig. \ref{fig:EMK} considering
Skyrme and relativistic EDFs. But still, the correlation among the
EDF families (Skyrme, Relativistic) are ordered. 

The integral in Eq. (\ref{eq:knucldef}) is taken between $\rho_0$/2 
and $\rho_0$, which is adapted to the linear regime around $\rho_c$
and corresponds to the typical dispersion of the density values around
the mean density in nuclei \cite{kha12}. Injecting expression
(\ref{eq:mlin}) in (\ref{eq:knucldef}) and assuming a Fermi shape of
the nuclear density, with diffusivity a $\simeq 0.5$~fm~\cite{rin80},
yield the analytical relation between the centroid of the GMR and $M_c$,
using Eq. (\ref{eq:egmrka}) and (\ref{eq:kaexp}):

\begin{widetext}
\begin{equation}
E_{\rm
GMR}=\frac{\hbar}{R}\left\{\frac{20\pi}{3mA}\int_{\rho_0/2}^{\rho_0}
\left[a\ln\left(\frac{\rho_0}{\rho}-1\right)+R\right]^2
\left(\frac{M_c\rho}{3\rho_c}-\frac{M_c}{3}+ K_c\right) 
\frac{a}{1-\rho/\rho_0}\frac{\rho_0^2}{\rho^2}\mathrm{d}\rho
+\frac{5K_\mathrm{Coul}}{3m}Z^2 A^{-4/3}\right\}^{1/2}
\label{eq:katop}
\end{equation}
\end{widetext}

The integral in Eq. (\ref{eq:katop}) denotes the nuclear contribution
whereas the second part comes from the Coulomb effects. The Coulomb
contribution is evaluated using K$_\mathrm{Coul}$=-5.2 MeV
\cite{bla80,sag07}. This value is obtained from the liquid drop
expansion of the incompressibility and applied to several Skyrme
interactions \cite{sag07}. It should be noted that the Fermi shape is
a good approximation of the density and we have checked that the
diffusivity of the density obtained from microscopic Hartree-Fock
calculations (0.47 fm) is very close to 0.5 fm. The use of the Fermi
density is legitimized by the aim of tracing the analytical impact of
the quantity $M_c$ on the GMR centroid. Equation (\ref{eq:katop})
also underlines the important role of the quantity $M_c$ on the GMR
centroid. On the contrary, the incompressibility at saturation density
$K_\infty$ doesn't play any specific role in Eq. (\ref{eq:katop}).  It
is therefore rather the quantity $M_c$ which is the relevant
quantity to be constrained by the GMR measurements.

\begin{figure}[tb]
\begin{center}
\scalebox{0.35}{\includegraphics{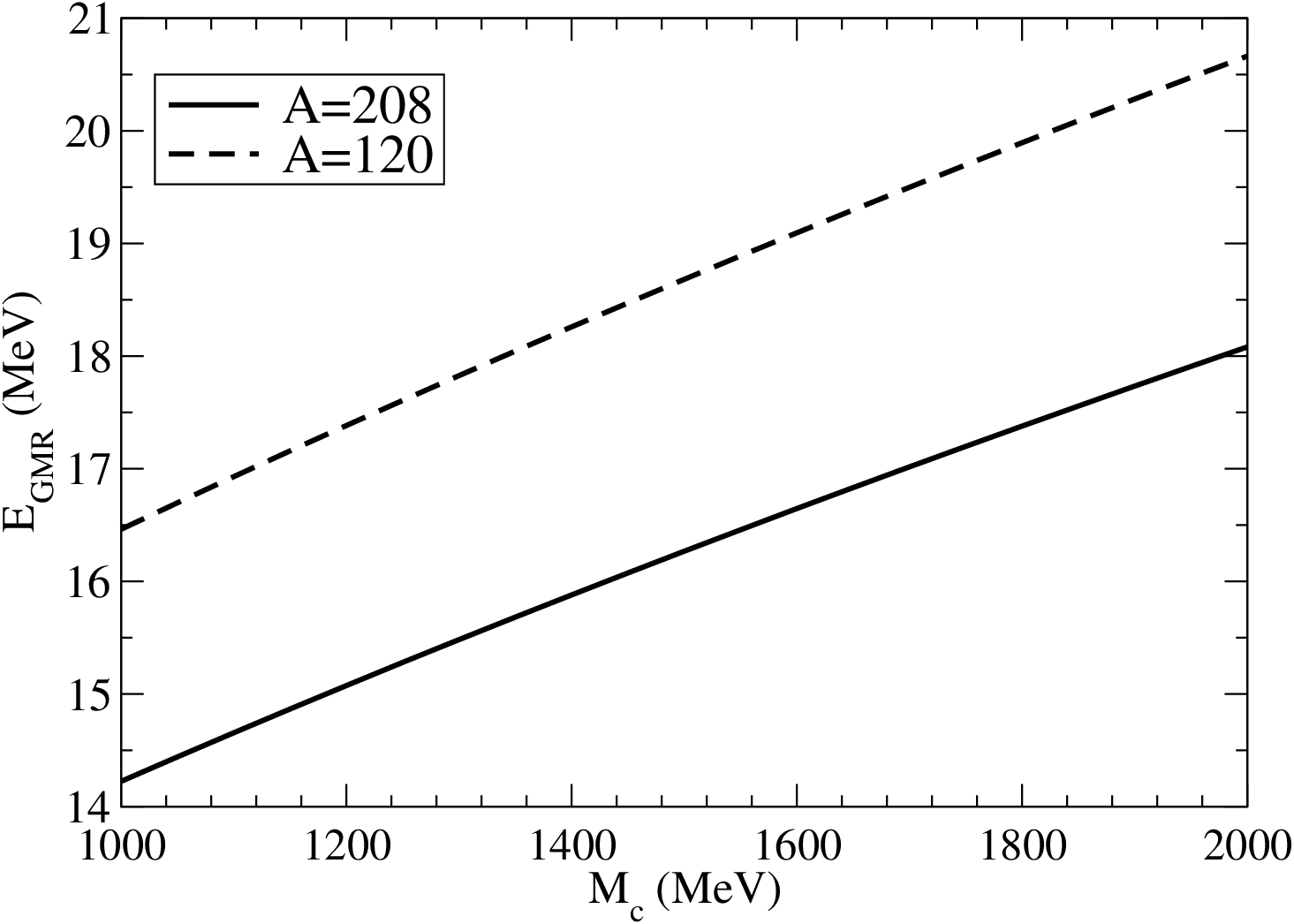}} \caption{Centroid of the
GMR in A=208 (solid line) and A=120 (dashed line) nuclei calculated
with the local density approximation for the nuclear incompressibility
and using its first derivative at the crossing point (Eq.
(\ref{eq:katop}) without the Coulomb term).} \label{fig:Mgmr}
\end{center}
\end{figure}

\subsection{Results and comparison with the microscopic method}

The stability of the results obtained with Eq.
(\ref{eq:katop}) has been studied with respect to the diffusivity
value $a$, the LDA prescription (Eq. (\ref{eq:knucldef})), the crossing
point ($\rho_c$,K$_c$) values and the integration range. A sound
stability is obtained against these quantities: the predicted GMR
centroid doesn't change by more than 10 \% by making all these
variations in relevant physical ranges.

We first study the behavior of the nuclear contribution
($K_\mathrm{Coul}$=0 in Eq. (\ref{eq:katop})). Fig. \ref{fig:Mgmr}
displays the correlation between the centroid of the GMR and the $M_c$
value using Eq. (\ref{eq:katop}), for nuclei with A=208 and A=120. A
good qualitative agreement is obtained with the fully microscopic
results (see Fig. \ref{fig:EMK} and \ref{fig:EMKSn}) in view of the
approximations performed to derive Eq. (\ref{eq:katop}). The
$A$-dependence is also well described. These results confirm the
validity of the present approach, and emphasize $M_c$ as a relevant
EoS quantity to be constrained by the GMR measurements. It also
qualitatively agrees with the microscopic results.

\begin{figure}[tb]
\begin{center}
\scalebox{0.35}{\includegraphics{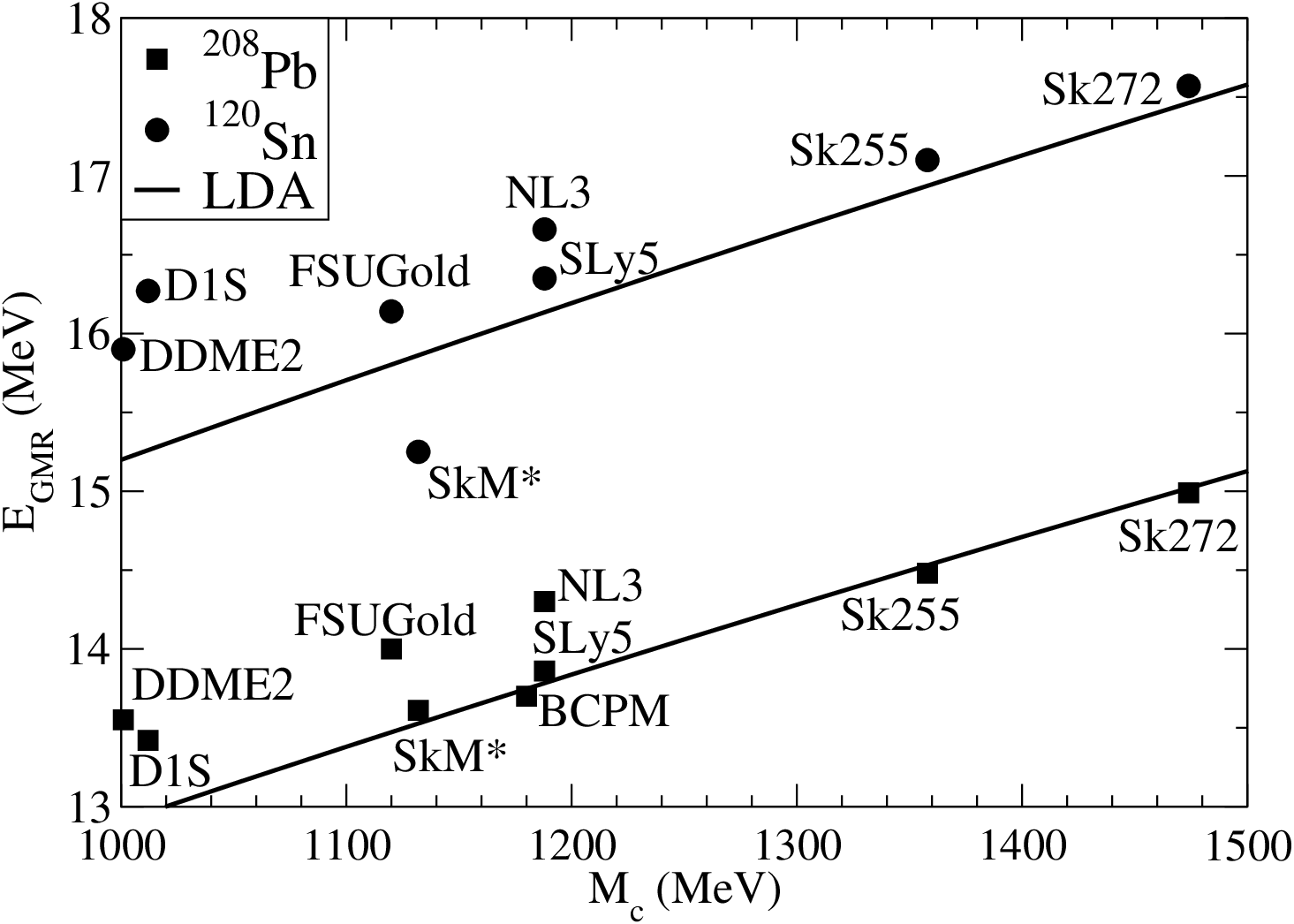}}
\caption{Centroid of the GMR in $^{208}$Pb and $^{120}$Sn calculated
with the LDA and including the Coulomb effects (solid lines, see text)
as a function of $M_c$. The values for various EDF obtained
microscopically (squares and dots) are also displayed for comparison.}
\label{fig:Mgmrtot}
\end{center}
\end{figure}

In order to perform a more quantitative study, the curves on 
Fig.~\ref{fig:Mgmrtot} displays the centroid of the GMR in $^{208}$Pb and
$^{120}$Sn nuclei predicted using the full Eq. (\ref{eq:katop}) with both the
nuclear and the Coulomb contributions. The comparison with the
microscopic results using various EDFs is also shown. A good agreement
is obtained in both cases, strengthening again the present analytical
LDA approach, and emphasising the role of $M_c$. Comparing Figs.
\ref{fig:Mgmr} and \ref{fig:Mgmrtot}, the Coulomb effect on the GMR
centroid can be evaluated to be about 1 MeV in heavy nuclei.

The almost linear correlation between $E_\mathrm{GMR}$ and $M_c$ observed on Fig. \ref{fig:Mgmrtot} 
can be further investigated. Eq. (\ref{eq:katop}) can be rewritten as:

\begin{equation}
E_{GMR}=\big(\alpha(A,\rho_0)M_c+\beta(A,Z,\rho_0)\big)^{1/2}
\label{eq:elin}
\end{equation}

with 
\begin{eqnarray}
\alpha(A,\rho_0)&\equiv&\frac{20\pi\hbar^2}{9mAR^2}\int_{\rho_{min}}^{\rho_0}
\left[a\ln\left(\frac{\rho_0}{\rho}-1\right)+R\right]^2 \times \nonumber \\
&&\hspace{1cm}\left(\frac{\rho}{\rho_c}-1\right) 
\frac{a}{1-\rho/\rho_0}\frac{\rho_0^2}{\rho^2}\mathrm{d}\rho
\label{eq:alpha}
\end{eqnarray}

\begin{eqnarray}
\beta(A,Z,\rho_0)&\equiv&\frac{5\hbar^2K_\mathrm{Coul}}{3mR^2}Z^2 A^{-4/3} +
\nonumber \\
&&\hspace{-1.8cm}\frac{20\pi\hbar^2}{3mAR^2}\int_{\rho_{min}}^{\rho_0}
\left[a\ln\left(\frac{\rho_0}{\rho}-1\right)+R\right]^2
\frac{aK_c}{1-\rho/\rho_0}\frac{\rho_0^2}{\rho^2}\mathrm{d}\rho \nonumber \\
\label{eq:beta}
\end{eqnarray}
Fixing $\rho_c$ and $K_c$, the coefficients $\alpha$ and
$\beta$ only depend on the nucleus' mass and
charge ($A$, $Z$), and on the saturation density $\rho_0$, which is
constrained by the charge radii. Typical values are $\alpha$=0.12 MeV 
and $\beta$=42 MeV$^2$ in the case of $^{208}$Pb and $\alpha$=0.16 MeV 
and $\beta$=75 MeV$^2$ in the case of $^{120}$Sn.

It should be noted that the LDA approximation (\ref{eq:elin}) of
E$_{GMR}$ can be obtained because of the existence of the crossing
point. In Eq.~(\ref{eq:elin}), the energy of the GMR depends on the
functional mostly through the parameter $M_c$. In conclusion, the LDA
allows to obtain expression~(\ref{eq:elin}) relating $E_\mathrm{GMR}$
with $M_c$ in a simple and accurate form.

Introducing ($M_0$,$E_0$) as the reference point, where $M_0\equiv$1200~MeV 
and $E_0$ is the corresponding GMR energy, one can go one step further and
linearise Eq. (\ref{eq:elin}) with respect to $M_c-M_0$, as
\begin{equation}
E_{GMR}\simeq \frac{\alpha}{2E_0} M_c+\left(E_0-\frac{\alpha
M_0}{2E_0}\right)
\label{eq:elin2}
\end{equation}
This is justified for the typical values of M$_c$, ranging between 1000~MeV and 1500~MeV 
as shown on Fig. \ref{fig:Mgmrtot}.
The almost linear correlation between
$E_{GMR}$ and $M_c$ observed on Fig. \ref{fig:Mgmrtot} is therefore understood
by the present approach (Eq. (\ref{eq:elin2})). It clearly shows that the
measurement of the GMR position constrains $M_c$, which is a first
information on the density dependence of the incompressibility. It
should be recalled that such a quantitative description is not
possible with $K_\infty$ because there is no crossing point of the
incompressibility at saturation density: Eq. (\ref{eq:katop}) is not
applicable in that case.

\section{Conclusions}
\label{sec:conclusion}

The relationship between the isoscalar GMR and the equation of state
raises the question of which EoS quantity is constrained by GMR
centroid measurements. The incompressibility modulus $K_\infty$ 
alone may not the relevant one nor
the most direct because the more general density dependence of the
incompressibility should be considered. A crossing area is observed on
$K(\rho)$ at $\rho_c\simeq 0.1$~fm$^{-3}$ among various functionnals.
Using a microscopic approach, such as constrained-HFB, the
slope $M_c$ of $K(\rho)$ at the crossing density can be directly
constrained by GMR measurements. This shall assess the change of the
method in extracting EoS quantities from GMR: $M_c$ is first
constrained, and an approximate value of $K_\infty$ can be deduced in
a second step \cite{kha12}. 

The stability of $K_c$ has been demonstrated in the case of Skyrme
EDFs. A general relationship between $M_c$ and $K_\infty$ is obtained,
showing the contribution of the uncertainty in the density
dependence of the EoS which has been casted into the quantities
$Q_\infty$. The $K_\infty$ value can be determined in a second step
from the knowledge of the $M_c$ value, requiring a better constraint
on the skewness parameter $Q_\infty$, being the main uncertainty for the
density dependence of the incompressibility between the
crossing density and the saturation density. 
One should recall that
the $K_\infty$ value remains 230 $\pm$ 40 MeV (17\% uncertainty),
whereas the quantity $M_c$ is better constrained to be $M_c$=1100
$\pm$ 70 MeV (6\% uncertainty) \cite{kha12}. A better knowledge of
higher order density dependent terms of $E/A(\rho)$, e.g. the skewness
parameter $Q_\infty$, shall help to more
accurately relate the parameter $M_c$ to the incompressibility
modulus $K_\infty$.

Using the LDA approach and an analytical approximation of the density,
the microscopic results have been confirmed: the measurement of the
centroid of the isoscalar giant monopole resonance constrains the
first derivative $M_c$ of the incompressibility around the crossing
point $\rho_c\simeq 0.1$~fm$^{-3}$.  A analytical relation between the
centroid of the GMR and the quantity $M_c$ is derived and the
predicted GMR centroid are found in good agreement with the
microscopic method.

\section*{Acknowledgement}

The authors thank I. Vida\~na for fruitful discussions. This work has
been partly supported by the ANR SN2NS contract, the Institut
Universitaire de France, and by CompStar, a Research Networking
Programme of the European Science Foundation.

\end{document}